\documentclass[11pt]{article}

\usepackage{fullpage}
\usepackage{amssymb,amsmath}
\usepackage{graphicx}
\usepackage{setspace}

\title{Re-Pair Compression of Inverted Lists}
\author{Francisco Claude\footnote{David R. Cheriton School of Computer Science, University of Waterloo, Canada. \texttt{fclaude@cs.uwaterloo.ca}} \and
Antonio Fari\~na\footnote{Department of Computer Science, University of A Coru\~na, Spain. \texttt{fari@udc.es}} \and
Gonzalo Navarro\footnote{Department of Computer Science, University of Chile, Chile. \texttt{gnavarro@dcc.uchile.cl}} }

\newtheorem{obs}{Observation}
\newtheorem{theorem}{Theorem}
\newtheorem{coro}{Corollary}
\newcommand{\bss}{\mathcal{S}}
\newcommand{\C}{\mathcal{C}}
\newcommand{\svs}{{\em svs}}
\newcommand{\bys}{{\em by}}
\newcommand{\merge}{{\em merge}}

\newcommand{\expon}{{\em exp}}
\newcommand{\lookup}{{\em lookup}}
\newcommand{\repair}{{\em Re-Pair}}
\newcommand{\rice}{{\em Rice}}

\begin{document} 

\maketitle 

\begin{abstract}
Compression of inverted lists with methods that support fast intersection
operations is an active research topic. Most compression schemes rely on
encoding differences between consecutive positions with techniques that
favor small numbers. In this paper we explore a completely different
alternative: We use \repair\ compression of those differences. While \repair\ by
itself offers fast decompression at arbitrary positions in main and secondary
memory, we introduce variants that in addition speed up the operations required
for inverted list intersection. We compare the resulting data structures with
several recent proposals under various list intersection algorithms, to
conclude that our \repair\ variants offer an
interesting time/space tradeoff for this problem, yet further improvements
are required for it to improve upon the state of the art. 
\end{abstract}

\section{Introduction}

Inverted indexes are one of the oldest and simplest data structures ever
invented, and at the same time one of the most successful ones in Information
Retrieval (IR) for natural language text collections. They play a central role
in any book on the topic \cite{BYRN99,WMB99}, and are also
at the heart of most modern Web search engines, where simplicity is a plus.

An inverted index is a vector of lists. Each vector entry
corresponds to a different {\em word} or {\em term}, and its list points to
the {\em occurrences} of that word in the text collection. The collection is
seen as a set of {\em documents}. The set of different words is called the
{\em vocabulary}. Empirical laws well accepted in IR \cite{Hea78} establish
that the vocabulary is much smaller than the collection size $n$, more precisely
of size $O(n^\beta)$, for some constant $0<\beta<1$ that depends on the text
type.

Two main variants of inverted indexes exist \cite{BYMN02,ZM06}. One is aimed at
retrieving documents which are ``relevant'' to a query, under some criterion.
Documents are regarded as vectors, where terms are the dimensions, and the
values of the vectors correspond to the relevance of the terms in the
documents. The lists point to the documents where each term appears, storing
also the weight of the term in that document (i.e., the coordinate value). The
query is seen as a set of words, so that retrieval consists in processing the
lists of the query words, finding the documents which, considering the weights
the query terms have in the document, are predicted to be relevant. Query
processing usually involves somehow merging the involved lists, so that
documents can get the combined weights over the different terms. Algorithms
for this type of queries have been intensively studied, as well as different
data organizations for this particular task \cite{PZSD96,WMB99,ZM06,AM06,SC07}.
List entries are usually sorted by descending relevance of the term in
the documents.

The second variant, which is our focus, are the inverted indexes
for so-called {\em full-text retrieval}. These are able of finding
the documents where the query appears. In this case the lists
point to the documents where each term appears, usually in
increasing document order. Queries can be single words, in which
case the retrieval consists simply of fetching the list of the
word; or disjunctive queries, where one has to fetch the
lists of all the query words and merge the sorted lists; or
conjunctive queries, where one has to intersect the lists.
While intersection can be done also by scanning all the lists in
synchronization, it is usually the case that some lists are much
shorter than the others \cite{Zip49}, and this opens many 
opportunities of faster intersection algorithms. Those are even
more relevant when many words have to be intersected.

Intersection queries have become extremely popular because of Google-like
default policies to handle multiword queries. Given the huge size of the
Web, Google solves the queries, in principle, by intersecting the 
inverted lists. Another important query where
intersection is essential is the phrase query. This can be solved by
intersecting the documents where the words appear and then postprocessing the
candidates. In order to support phrase queries at the index level, the inverted
index must store all the positions where each word appears in each document.
Then phrase queries can be solved essentially by intersecting word positions.
The same opportunities for smart intersection arise.

The amount of recent research on intersection of inverted lists witnesses the
importance of the problem
\cite{DM00,BK02,BY04,BYS05,BLOL06,ST07,CM07}. Needless to say,
space is an issue in inverted indexes,
especially if one has to store word positions. Much research has been carried
out on compressing inverted lists \cite{WMB99,NMNZBY00,ZM06,CM07}, and on its
interaction with the query algorithms, including list intersections. Despite
that algorithms in main memory have received much attention, in many cases one
resorts to secondary memory, which brings new elements to the tradeoffs.
Compression not only reduces space, but also transfers from secondary memory
when fetching inverted lists (the vocabulary usually fits in main memory even
for huge text collections). Yet, the random accesses used by the smart
intersection algorithms become more expensive.

Most of the list compression algorithms rely on the fact that the inverted lists
are increasing, and that the differences between consecutive entries are
smaller on the longer lists. Thus, a scheme that represents those differences
with encodings that favor small numbers work well \cite{WMB99}. Random access
is supported by storing sampled absolute values, and some storage schemes for
those samples considering secondary storage have been proposed as well
\cite{CM07}.

In this paper we explore a completely different compression method for
inverted lists: We use \repair\ compression of the differences. \repair\
\cite{LM00} is a grammar-based compression method that generates a dictionary
of common sequences, and then represents the data as a sequence of those
dictionary entries. It is simple and fast at decompression, allowing for
efficient random access to the data even on secondary memory, and achieves
competitive compression ratios. It has been successfully used for compression
of Web graphs \cite{CN07}, where the adjacency lists play the role of inverted
lists. Using it for inverted lists is more challenging because several other
operations must be supported apart from fetching a list, in particular the
many algorithms for list intersection. We show that \repair\ is well-suited for
this task as well, and also design new variants especially tailored to the
various intersection algorithms.
We test our techniques against a number of the best existing compression
methods and intersection algorithms, showing that our \repair\ variants offer an
interesting time/space tradeoff for this problem, yet further improvements
are required for it to improve upon the state of the art.

\section{Related Work}
\label{sec:related}

\subsection{Intersection algorithms for inverted lists}
\label{sec:iialg}

The intersection of two inverted
lists can be done in a merge-wise fashion (which is the best
choice if both lists are of similar length), or using a
set-versus-set (\svs) approach where the longer list is searched for
each of the elements of the shortest, to check if they should
appear in the result. Either binary or exponential (also called galloping or
doubling) search are
typically used for such task. The latter checks the list at positions
$i+2^j$ for increasing $j$, to find an element known to be after position
$i$ (but probably close). All these approaches assume that the lists
to be intersected are given in sorted order.

Algorithm \bys~\cite{BY04} is based on binary searching the longer list
$N$ for the median of the smallest list $M$. If the median is found, it is
added to the result set. Then the algorithm proceeds recursively on the left
and right parts of each list. At each new step the longest sublist is
searched for the median of the shortest sublist. Results showed that \bys\
performs about the same number of comparisons than \svs\ with binary search.
As expected, both \svs\ and \bys\ improve \merge\ algorithm when 
$|N|>>|M|$ (actually from $|N| \approx 20 |M|$).

Multiple lists can be intersected using any pairwise \svs\
approach (iteratively intersecting the two shortest lists, and then the
result against the next shortest one, and so on). Other algorithms are
based on choosing the first element of the smallest list as an {\em eliminator}
that is searched for in the other lists (usually keeping track of the
position where the search ended). If the eliminator
is found, it becomes a part of the result. In any case,
a new eliminator is chosen. Barbay et al. \cite{BLOL06} compared four multi-set
intersection algorithms: {\em i)} a pairwise \svs-based
algorithm; {\em ii)} an eliminator-based algorithm \cite{BK02} (called
{\em Sequential}) that chooses the eliminator
cyclically among all the lists and exponentially searches for
it; {\em iii)} a multi-set version of \bys; and {\em iv)} a
hybrid algorithm (called {\em small-adaptive}) based on \svs\ and on
the so-called {\em adaptive algorithm} \cite{DM00}, which at each step
recomputes the list ordering according to their elements not yet processed,
chooses the eliminator from the shortest list, and tries the others in
order.
Results \cite{BLOL06} showed that the
simplest pairwise \svs-based approach (coupled with exponential search)
performed best.

\subsection{Data structures for inverted lists}
\label{sec:iids}

The previous algorithms require that lists can be accessed at any
given element (for example those using binary or exponential search)
and/or that, given a value, its smallest successor from a list can
be obtained. Those needs interact with the inverted list compression
techniques.

The compression of inverted lists usually represents each list
$\langle p_1,p_2,p_3,\ldots,p_\ell\rangle$ as a sequence of d-gaps $\langle
p_1,p_2-p_1, p_3-p_2,\ldots,p_\ell-p_{\ell-1}\rangle$, and uses
a variable-length encoding for these differences, for example
$\gamma$-codes, $\delta$-codes, Golomb codes, etc.~\cite{WMB99}.
More recent proposals \cite{CM07} use byte-aligned codes, which lose
little compression and are faster at decoding.

Intersection of compressed inverted lists is still possible using
a merge-type algorithm. However, approaches that require
direct access are not possible as sequential decoding of the d-gaps
values is mandatory. This problem can be overcome by sampling
the sequence of codes \cite{CM07,ST07}. The result is
a two-level structure composed of a top-level array
storing the absolute values of, and pointers to, the sampled values in the
sequence, and the encoded sequence itself.

Assuming $1\le p_1< p_2<\ldots<p_\ell\le u$,
Culpepper and Moffat \cite{CM07} extract a sample every $k' = k \log \ell$
values\footnote{Our logarithms are in base 2 unless otherwise stated.}
from the compressed list, being $k$ a parameter. Each of those samples
and its corresponding offset in the compressed sequence is
stored in the top-level array of pairs $\langle value,of\!fset
\rangle$ needing $\lceil \log u \rceil$ and
$\lceil \log(\ell\log(u/\ell))\rceil$ bits,
respectively, while retaining random access to the
top-level array. Accessing the $v$-th value of the compressed
structure implies accessing the sample $\lceil v/k'\rceil$
and decoding at most $k'$ codes.
We call this ``(a)-sampling''.
Results showed that intersection using \svs\ coupled with
exponential search in the samples performs just slightly worse
than \svs\ over uncompressed lists.

Sanders and Transier \cite{ST07}, instead of sampling at regular intervals of
the list, propose sampling regularly at the domain values. We call this
a ``(b)-sampling''. The
idea is to create buckets of values identified by their most
significant bits and building a top-level array of pointers to
them. Given a parameter $B$ (typically $B=8$), and the value $k =
\lceil \log (uB/\ell) \rceil$, bucket $b_i$  stores the values $x_j =
p_j ~\textrm{mod}~ 2^{k}$ such that $(i-1)2^k \leq p_j < i\,2^k$.
Values $x_j$  can also be compressed (typically using variable-length
encoding of d-gaps). Comparing with the previous approach \cite{CM07},
this structure keeps only pointers in the top-level array, and
avoids the need of searching it (in sequential, binary, or exponential
fashion), as $\lceil p_j/2^k\rceil$
indicates the bucket where $p_j$ appears. In exchange, the blocks are of
varying length and more values might have to be scanned on average for
a given number of samples. The authors also
keep track of up to where they have decompressed
a given block in order to avoid repeated decompressions.
This direct-access method is called \lookup.

Moffat and Culppeper \cite{CulpepperM_adcs07} proposed a technique
to further improve the space and time of inverted lists
representation. The main idea is to represent longer lists using
bitmaps. Since the longer lists generate very dense bitmaps, the
next element can be retrieved in low amortized time, and also the
intersection between two long lists can be done by bit-{\sc and}
operations. Experimental results show that this effectively improves
the speed and achieves lower overall space. As we show in our 
experimental results, their
technique can be applied to our approach, yet it
does not yield to such an effective improvement in this case.

\subsection{Re-Pair compression algorithm}
\label{sec:repair}

\repair\ \cite{LM00} consists of repeatedly
finding the most frequent pair of symbols in a sequence of integers and
replacing it with a new symbol, until no more replacements are useful.
More precisely, \repair\ over a sequence $L$ works as follows:
\begin{enumerate}
\item It identifies the most frequent pair $ab$ in $L$.
\item It adds the rule $s\rightarrow ab$ to a dictionary $R$, where $s$ is a
new symbol not appearing in $L$.
\item It replaces every occurrence of $ab$ in $L$ by $s$.\footnote{As far as
possible, e.g., one cannot replace both occurrences of $aa$ in $aaa$.}
\item It iterates until every pair in $L$ appears once.
\end{enumerate}

We call $C$ the sequence resulting from $L$ after compression. Every symbol in
$C$ represents a {\em phrase} (a substring of $L$), which is of length $1$
if it is an original symbol (called a {\em terminal}) or longer if it is an
introduced one (a {\em non-terminal}). Any phrase can be recursively expanded
in optimal time (that is, proportional to its length), even if $C$ is stored on
secondary memory (as long as the dictionary $R$ fits in RAM). Notice that replaced
pairs can contain terminal and/or nonterminal symbols.

\repair\ can be implemented in linear time \cite{LM00}. However, this requires
several data structures to track the pairs that must be replaced. This is
problematic when applying it to large sequences, as witnessed when using it 
for compressing natural language text \cite{Wan03}, suffix arrays \cite{GN07}, 
and Web graphs \cite{CN07}. The space consumption of the linear time algorithm
is about $5|L|$ words.

In this work, we make use of an approximate version \cite{CN07} that provides a 
tuning parameter that trades speed and memory usage for compression ratio. It 
achieves very good compression ratio within reasonable time and using little memory 
(3\% on top of the sequence). It also works well on secondary memory.
The main ideas behind this approximate version are to replace 
many pairs per iteration and to count the occurrences of 
each pair using limited-capacity hash tables, so that only the pairs occurring early 
in $L$ are considered. As the compression method advances, the space reduced 
from the original sequence is added to the space reserved for the hash tables. 
This extra space improves the selection of pairs at the later iterations of 
the algorithm, when the distribution of the occurrences of the pairs is expected to be 
flatter, and hence more precision is needed to choose good pairs.

Larsson and Moffat \cite{LM00} proposed a method to compress the set of rules
$R$. In this work we prefer another method \cite{GN07}, which is not so
effective but allows accessing any rule without decompressing the whole set of
rules. It represents the DAG of rules as a set of trees. Each tree
is represented as a sequence of leaf values (collected into a sequence $R_S$)
and a bitmap that defines the tree shapes in preorder (collected into a bitmap
$R_B$). Nonterminals are represented by the starting position of their tree (or
subtree) in $R_B$. In $R_B$, internal nodes are represented by 1s and leaves
by 0s, so that the value of the leaf at position $i$ in $R_B$ is found at
$R_S[rank_0(R_B,i)]$. Operation $rank_0$ counts the number of 0s in $R_B[1,i]$
and can be implemented in constant time, after a linear-time preprocessing that
stores only $o(|R_B|)$ bits of space
\cite{Mun96} on top of the bitmap. To expand a nonterminal, we traverse $R_B$ and extract the leaf
values, until we have seen more 0s than 1s. Leaf values corresponding to
nonterminals must be recursively expanded. Nonterminals are shifted by the
maximum terminal value to distinguish them.

\begin{figure*}[htbp]
\begin{center}
\includegraphics[width=0.8\textwidth]{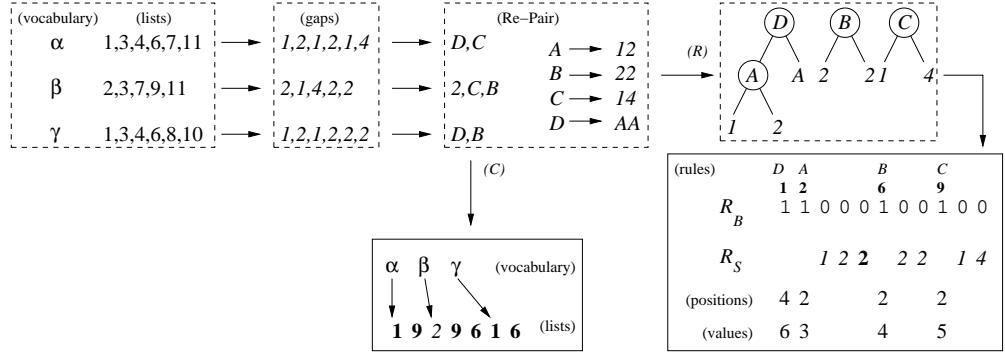}
\caption{Example of inverted lists compressed with our \repair-based method.
Solid boxes enclose the data we represent. We include both variants of data
aligned to the 1s in $R_B$. Bold numbers (nonterminals) in the lists
and $R_S$ refer to positions in $R_B$, whereas slanted ones (terminals)
refer to gap values. To distinguish them, the maximum offset $u$ is actually
added to the bold numbers.}
\label{fig:example}
\end{center}
\end{figure*}

Figure~\ref{fig:example} shows an example. Consider the second box (gaps) as
the text to be compressed. Its most frequent pair is $\langle \mathit{1},
\mathit{2}\rangle$. Hence we add rule $A \rightarrow \mathit{1\,2}$ to the
dictionary $R$ and replace all the occurrences of $\mathit{1\,2}$ by nonterminal
$A$. We go on replacing pairs; note that the fourth rule $D \rightarrow AA$
replaces nonterminals. In the final sequence $D\,C\,\mathit{2}\,C\,B\,D\,B$,
no repeated pair appears. We now represent the dictionary of four rules as
a forest of four small subtrees. Now, as nonterminal $A$ is used in the
right-hand side of another rule, we insert its tree as a subtree of one such
occurrence, replacing the leaf. Other occurrences of $A$ are kept as is (see
leftmost box in the first row). This will save one integer in the
representation of $A$. The final representation is shown in the large box below
it. In $R_B$, the shape of the first subtree (rooted at $D$) is represented by
$11000$ (the first 1 corresponds to $D$ and the second to $A$); the other two
($B$ and $C$) are $100$. These nonterminals will be further identified by the
position of their 1 in $R_B$: $D=\mathbf{1}$, $A=\mathbf{2}$, $B=\mathbf{6}$,
$C=\mathbf{9}$. Each 0 (tree leaf) corresponds to an entry in $R_S$, containing
the leaf values: $\mathit{12}A = \mathit{12}\mathbf{2}$ for the first subtree,
and $\mathit{22}$ and $\mathit{14}$ for the others. Nonterminal positions (in
boldface) are in practice distinguished from terminal values (in italics) by
adding them the largest terminal value. Finally, sequence $C$ is
$\mathbf{1\,9}\,\mathit{2}\,\mathbf{9\,6\,1\,6}$. To expand, say, its sixth
position ($C[6]=\mathbf{1}$), we scan from $R_B[1,\ldots]$ until we see more 0s
than 1s, i.e., $R_B[1,5] = 11000$. Hence we have three leaves, namely the first
three 0s of $R_B$, thus they correspond to the first three positions of
$R_S$, $R_S[1,3] = \mathit{12}\mathbf{2}$. Whereas $\mathit{12}$ is already
final (i.e., terminals), we still have to recursively expand $\mathbf{2}$.
This corresponds to subtree $R_B[2,4] = 100$, that is, the first and second 0
of $R_B$, and thus to $R_S[1,2] = \mathit{12}$. Concatenating, $C[6]$ expands
to $\mathit{1212}$.

\section{Our Compressed Inverted Index}

\subsection{Re-Pair compressed inverted lists}

Our basic idea is to differentially encode the inverted lists, transforming a
sequence $\langle p_1,p_2,p_3,\ldots,p_\ell\rangle$ into $\langle p_1,p_2-p_1,$
$p_3-p_2,\ldots,p_\ell-p_{\ell-1}\rangle$, and then apply the \repair\
compression
algorithm to the sequence formed by the concatenation of all the lists. A
unique integer will be appended to the beginning of each list prior to the
concatenation, in order to ensure that no \repair\ phrase will span more than
one list. At the end of the compression process, we remove those artificial
identifiers from the compressed sequence $C$ of integers. We store
a pointer from each vocabulary entry to the first integer of $C$ that
corresponds to its inverted list. We must also store the \repair\ dictionary,
as explained in Section~\ref{sec:repair}.
The terminal symbols are directly the corresponding differential values,
e.g., value $\mathit{3}$ is represented by the terminal integer value 3.

Since no phrase extends over two lists, any list can be expanded in
optimal time (i.e., proportional to its uncompressed size), by expanding all
the symbols of $C$ from the corresponding vocabulary pointer to the next.
Moreover, if the dictionary is kept in main memory and
the compressed lists on disk, then the retrieval accesses at most
$1+\lceil (\tilde{\ell}-1)/B\rceil$ contiguous disk blocks, where $B$ is the
disk block size and $\tilde{\ell} \le \ell$ is the length of the compressed
list. Thus I/O time is also optimal in this sense.

\subsection{Direct access: sampling and skipping}
\label{sec:skipping}

Several intersection algorithms, as explained in Section~\ref{sec:iialg},
require direct accesses to the inverted lists.
With \repair\ compression there is no direct access to every list position,
even if we knew its absolute value and position in the compressed data. We
can have direct access only to the \repair\ phrase beginnings, that is, to
integers in the compressed sequence. Accepting those ``imprecisions'' at
locating, we can still implement the (a)- and (b)-samplings of 
Section~\ref{sec:iids}.

Before discussing sampling we note that,
unlike other compression methods, we can apply
some skipping without sampling nor decompressing, by processing the compressed
list symbol by symbol without expanding them. The key idea is that {\em
nonterminals also represent differential values}, namely the sum of the
differences they expand into. We call this the {\em phrase sum} of the
nonterminal. In our example, as $D = \mathbf{1}$ expands to $\mathit{1212}$, its
phrase sum is $1+2+1+2 = 6$. If we store this sum associated to $D$, we can
skip it in the lists without expanding it, by knowing its symbols add up to 6.

Phrase sums will be stored in sequence $R_S$, aligned with the 1 bits of
sequence $R_B$. Thus $rank$ is not anymore necessary to move from one sequence
to the other. The 0s in $R_B$ are aligned in $R_S$ to the leaf data, and the 1s
to the phrase sums of the corresponding nonterminals.

In order to find whether a given document $d$ is in the compressed list, we
first scan the entries in $C$, adding up in a sum $s$ the value $C[i]$ if it
is a terminal, or $R_S[C[i]]$ if it is a nonterminal. If at some point we get
$s=d$, then $d$ is in the list. If instead $s>d$ at some point, we
consider whether the last $C[i]$ processed is a terminal or not. If it is a
terminal, then $d$ is not in the list. If it is a nonterminal, we restart the
process from $s-C[i]$ and process the $R_S$ values corresponding to the 0s
in $R_B[C[i],\ldots]$, recursing as necessary until we get $s=d$ or $s>d$
after reading a terminal.

In our example of Figure~\ref{fig:example}, assume we want to know whether
document 9 is in the list of word $\beta$. We scan its list $\mathit{2}\,C\,B =
\mathit{2}\,\mathbf{9\,6}$, from sum $s=0$. We process $\mathit{2}$, and since
it is a terminal we set $s = s+2 = 2$. Now we process $\mathbf{9}$, and since
it is a nonterminal, we set $s = s + R_S[9] = s+5 = 7$ (note the $5$ is
correct because $\mathbf{9} = C$ expands to $\mathit{1\,4}$). Now we process
$\mathbf{6}$, setting $s = s + R_S[6] = s+4 = 11$. We have exceeded $d=9$,
thus we restart from $s=7$ and now process the zeros in $R_B[6,\ldots] =
100\ldots$. The first 0 is at $R_B[7]$, and since $R_S[7] = \mathit{2}$ is a
terminal, we add $s = s +R_S[7] = 9$, concluding that $d=9$ is in the list.
The same process would have shown that $d=8$ was not in the list.

We return to sampling now.
Depending on whether we want to use strategies of type \svs\ or \lookup\ for
the search, we can add the corresponding sampling of absolute values to the
\repair\ compressed lists. For \svs\ we will sample $C$ at regular positions
(i.e., (a)-sampling),
and will store the absolute values preceding each sample. The pointers to $C$
are not necessary, as both the sampling and the length of the entries of
$C$ are regular. This is a plus compared to classical gap encoding
methods. Strategy \lookup\ will insert a new sample each time the absolute
value surpasses a new multiple of a sampling step (i.e., (b)-sampling). 
Now we need to store
pointers to $C$ (as in the original method) and also the absolute values
preceding each sample (unlike the original method). The reason is that the
value to sample may be inside a nonterminal of $C$, and we will be able to point
only to the (beginning of the) whole nonterminal in $C$. Indeed, several
consecutive sampled entries may point to the same position in $C$.

In our example, imagine we wish to do a (b)-sampling on list $\gamma$,
for $2^k=4$ (including the first element too). Then the 
samples should point at positions 1, 3, and 5, of the
original list. But this list is compressed into $D\,B$, so the first two
pointers point to $D$, and the latter to $B$. That is, the sampling array stores
$(0,1),\,(0,1),\,(6,2)$. Its third entry, e.g., means that the first element
$\ge 8$
is at its 2nd entry in its compressed list, and that we should start from value
6 when processing the differences. For example, if we wish to access the first
list value exceeding 4, we should start from $(0,1)$, that is, accumulate
differences from $D$, starting from value 0, until exceeding 4.

\subsection{Intersection algorithm}

We can implement any of the existing intersection algorithms on top
of our \repair\ compressed data structure. In this paper we will try several
versions. A first does skipping without any sampling, yet using the stored
phrase sums for the nonterminals. A second version uses (a)-sampling
and implements an \svs\ strategy with sequential, binary, or
exponential search in the samples \cite{CM07}, also profiting from skipping.
A third version uses instead a (b)-sampling, adapted to \repair\ as
explained above, and uses {\em lookup} search strategy \cite{ST07}, that is,
direct access to the correct sample thanks to the sampling method used.

To intersect several lists, we sort them in increasing order of their
{\em uncompressed} length (which we must store separately). Thus we
proceed iteratively, searching in step $i$ the list $i+1$ for the elements of
the {\em candidate} list. In step 1 this list is (the uncompressed
form of) list 1, and in step $i>1$ it is the outcome of the
intersection at step $i-1$.

To carry out each intersection, the candidate list is sequentially traversed.
Let $x$ be its current element. We skip phrases of list $i+1$ (possibly aided
by the type of sampling chosen), accumulating
gaps until exceeding $x$, and then consider the previous and current cumulative
gaps, $x_1 \le x < x_2$. Then the last phrase represents the range $[x_1,x_2)$.
We advance in the shorter list until finding the largest $x' < x_2$. We will
process all the interval $[x,x']$ within the phrase representing $[x_1,x_2)$
by a recursive procedure: We expand nonterminals into their two components,
representing subintervals $[x_1,z)$ and $[z,x_2)$. We partition $[x,x']$
according to $z$ and proceed recursively within each subinterval, until we
reach an answer (a terminal) to output or the interval $[x,x']$ becomes empty.
Note, however, that our dictionary representation does not allow for this
recursive partition. We must instead traverse their sequence in $R_S$ and
add up gaps. Nonterminals found in $R_S$, however, can be skipped.

\subsection{Optimizing space}

The original \repair\ algorithm requires two integers of space per new rule
introduced, and thus, roughly, one should stop creating new symbols
for pairs that appear just twice in the sequence, as the two integers saved
in the sequence would be reintroduced in the dictionary. Reality is more complex
because of dictionary compression and the usage of the exact number of bits
to represent the integers. In our case, we must also consider the
cost of storing the phrase sums on the symbols we create.

We aim at finding the optimal point where to stop inserting new dictionary 
entries. We first complete the compression process (inserting all the entries
up to the end) and then successively unroll the last symbol added by \repair\,
so as to choose the value that minimizes the overall size. 

Let us call $\sigma$ the size of the alphabet in the original sequence.
Let $d=|R_S|$ be the number of elements in $R_S$, and $l=|R_B|$ the length of 
the bitmap in the dictionary measured in bits. The space required to represent 
each symbol in $C$ or $R_S$ is $\bss(l)=\lceil \log_2 (\sigma+l-2) \rceil$, so 
the total space is $(d+n)\bss(l)+l+o(l) = (d+n)\bss(l)+\C(l)$ bits. We call 
$\rho$ the overhead added to each rule (in units of $\bss(l)$ bits, as this
data is also in $R_S$); in our case $\rho=1$.

\begin{obs}
The last symbol added by \repair\, $s\rightarrow s_1s_2$, adds 
$\rho+c(s_1)+c(s_2)$ symbols in $R_S$, it reduces the space in $C$ by $k$ 
symbols if it occurs $k$ times, and requires $f(s)=1+c(s_1)+c(s_2)$ bits in 
$R_B$, where  $c(a)=1$ if rule $a$ is used by another rule previous to 
$s$ and $c(a)=0$ otherwise.  
\end{obs}

Hence we can predict the size of the final representation after expanding the last symbol.
So, by keeping the order 
in which pairs were added to the dictionary, their new value when the dictionary
is compressed, their frequency, and the number of elements referencing each 
rule, we can compute the optimal dictionary size in $O(d)$ time.
Then we must expand the dictionary symbols that are 
to be discarded, which costs time proportional to the size of the output.

\section{Analysis}

One can achieve worst-case time $O(m (1+\log \frac{n}{m}))$ to intersect two
lists of length $m < n$, for example
by binary searching the longer list for the median of the shortest and dividing
the problem into two \cite{BY04}, or by exponentially searching the longer
list for the consecutive elements of the shortest \cite{CM07}. This is a lower
bound in the comparison model, as one can encode any of the ${n \choose m}$
possible subsets of size $m$ of a universe of size $n$ via the results of the
comparisons of an intersection algorithm, so these must be
$\ge \log {n \choose m} \ge m\log\frac{n}{m}$ in the worst case, and the
output can be of size $m$. Better results are possible for particular classes
of instances \cite{BLOL06}.

We now analyze our skipping method, assuming that the derivation trees of our
rules have logarithmic depth (which is a reasonable assumption, as shown later
in the experiments, and also theoretically achievable \cite{Sak05}). We expand the 
shortest list if it is compressed, at $O(m)$ cost, and use skipping to find its
consecutive elements on the longer list, of length $n$ but compressed to
$n' \le n$ symbols by \repair. Thus, we pay $O(n')$ time for skipping over all
the phrases. Now, consider that we have to expand phrase $j$, of length $n_j$,
to find $m_j$ symbols of the shortest list, $\sum_{j=1}^{n'} n_j = n$,
$\sum_{j=1}^{n'} m_j = m$. Assume $m_j>0$ (the others are
absorbed in the $O(n')$ cost). In the worst case we will traverse all the nodes
of the derivation tree up to level $\log m_j$, and then carry out $m_j$
individual traversals from that level to the leaves, at depth $O(\log n_j)$.
In the first part, we pay $O(2^i\log\frac{m_j}{2^i})$ for the $2^i$ binary
searches within the corresponding subinterval $[x,x']$ of $m_j$ at that level
(an even partition of the $m_j$ elements into $m_j/2^i$ is the worst case),
for $0\le i \le \log m_j$. This adds up to $O(m_j)$. 
For the second part, we have $m_j$ individual searches for one element $x$,
which costs $O(m_j(\log n_j - \log m_j))$. All adds up to
$O(m_j (1+\log\frac{n_j}{m_j}))$. Added over all $j$, this is
$O(m(1+\log\frac{n}{m}))$, as the worst case is $n_j = \frac{n}{n'}$, $m_j =
\frac{m}{n'}$.

\begin{theorem}\label{thm:inter}
The intersection between two lists $L_1$ and $L_2$ of length $n$ and $m$ 
respectively, with $n>m$, can be computed in time 
$O\left(n' + m(1+\log\frac{n}{m})\right)$, where \repair\ compresses $L_1$ to $n'$ 
symbols using rules of depth $O(\log n)$.	
\end{theorem}

Theorem \ref{thm:inter} exposes
the need to use sampling to achieve the optimal worst-case
complexity. One absolute sample out of $\log\frac{n}{n'}$ phrases
in the lists would multiply the space by $1+\frac{1}{\log\frac{n}{n'}}$ (which
translates into a similar overall factor, in the worst case, when added over
all the inverted lists), and would reduce the $O(n')$ term to
$O(m\log\frac{n}{n'})$, which is absorbed by the optimal
complexity as this matters only when $m \le n'$. Recall also that the parse
tree traversal requires that we do not represent the \repair\ dictionary in
compressed form.

\begin{coro}
By paying $1+\frac{1}{\log\frac{n}{n'}}$ extra space factor, the intersection algorithm of Theorem \ref{thm:inter}
takes \linebreak $O\left(m(1+\log\frac{n}{m})\right)$ time. 
\end{coro}

\section{Experimental Results}

We focus on the incremental approach to solve intersections of sets of words,
that is, proceeding by pairwise intersection from the shortest to the longest
list, as in practice this is the most efficient approach \cite{BLOL06,CM07}.
Thus we measure the intersection of two lists. We
implemented the variants with and without sampling and,
based on previous experiments \cite{BLOL06,CM07,ST07}, we
compare with the following, most promising, basic techniques for list
intersections (more sophisticated methods build orthogonally on these):
\merge\ (the merging based approach), \expon\ (the \svs\ approach
with exponential search over the sampling of the longer list
\cite{CM07}; this was better than binary and sequential search, as expected),
and \lookup\ (\svs\ where the sampling is regular on the domain
and so the search on the samples is direct). We used byte codes
\cite{CM07} to encode the differential gaps in all of the competing approaches,
as this yields good time-space trade off.
\footnote{For different reasons (stability, publicness, uniformity,
etc.) the competing codes are not available, so we had to
reimplement all of them. For the final version we plan to leave all
our implementations public.} We also include the versions 
representing the longest lists using bitmaps \cite{CulpepperM_adcs07}. For
\repair, we extract the lists that would be represented by bitmaps
according to the technique, and then we proceed to the compression
phase. 

We measure {\sc Cpu} times in main memory. Our machine is an Intel Core 2 Duo T8300,
2.4GHz, 3MB cache, 4GB RAM, running Ubuntu 8.04 (kernel 2.6.24-23-generic).
We compiled with  \verb|g++| using \verb|-m32 -09| directives.

We have parsed collections FT91 to FT94 from TREC-4,
\footnote{\tt http://trec.nist.gov} of 519,569,227 bytes (or 495.50MB), into
its 210,138 documents (of about 2.4KB on average), and built the inverted
lists of its 502,259 different words (a word is a maximum string formed by
letters and digits, folded to lowercase), which add up to 50,285,802 entries.
This small-document scenario is the worst for our \repair\ index. We show also
a case with larger documents, by packing 10 of our documents into one;
here the inverted lists have 29,887,213 entries.

 We used \repair\ construction with parameter $k=10,000$
\cite{CN07}, which takes just $1.5$min to compress the whole collection.

\subsection{Space Usage}

\begin{figure}[htb]

\centerline{
\includegraphics[scale=0.45]{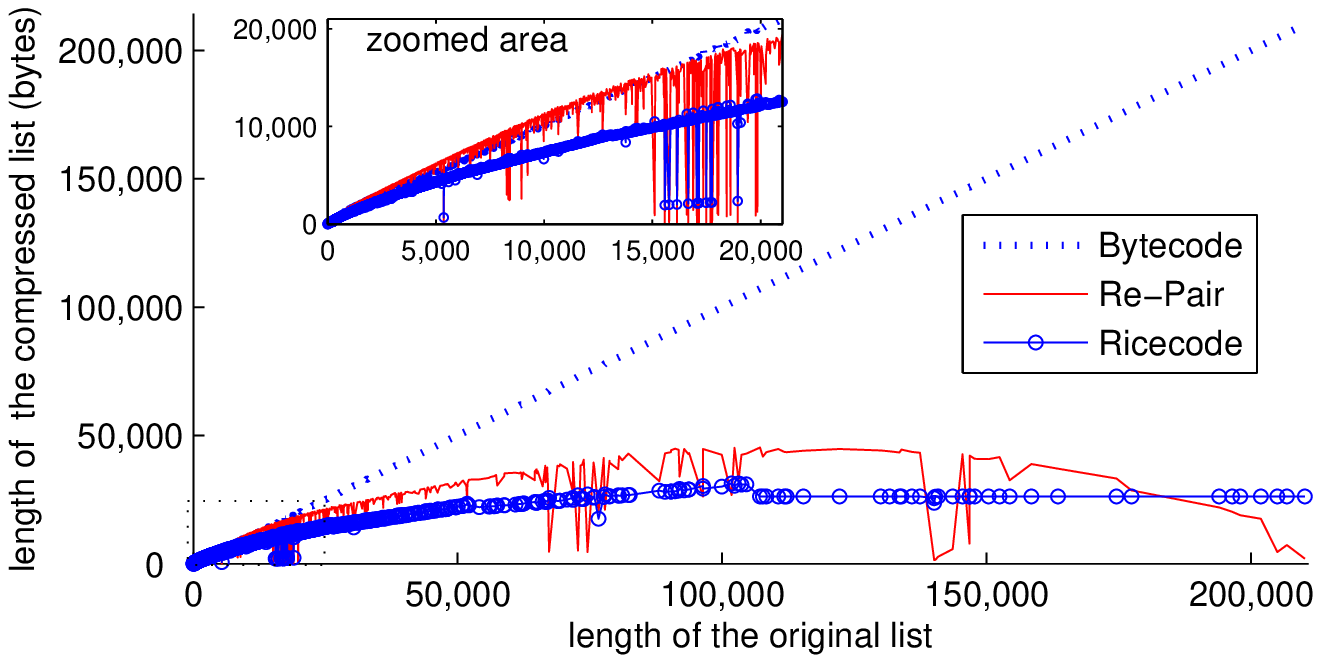}
\includegraphics[scale=0.45]{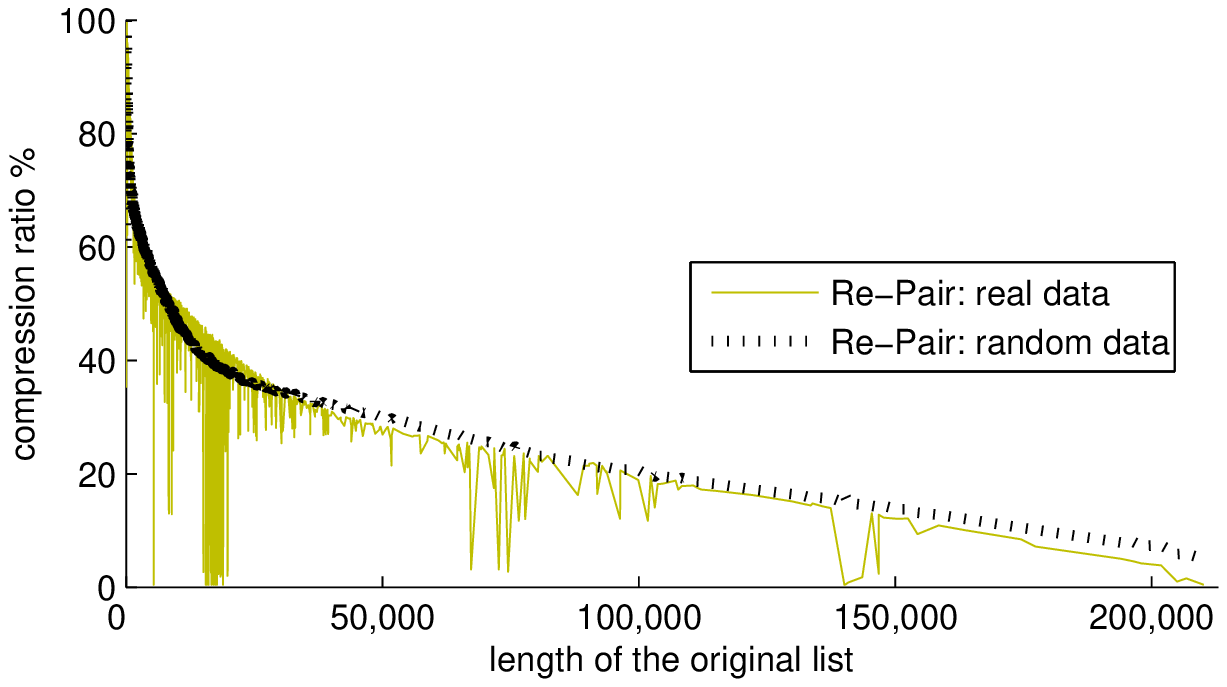}}

\caption{Left: The compressed list sizes as a function of their original length,
in bytes, dictionary excluded. Right: Compression ratio as a function of the length of the lists for random and real data.}
\label{fig:distr}
\end{figure}

\repair\ compression produces a non-monotonic phenomen\-on on the lengths of the
lists. Longer lists involve smaller and more repetitive differences, and
thus they compress much better (e.g. the pair of differences $(1,1)$ accounts
for around 10\% of the repetitions factored out by \repair).
Yet, expanding those resulting short lists is costly; this is why we process
the lists in the order given by their expanded length. Figure~\ref{fig:distr} (left) illustrates the resulting (non-monotonic) lengths.

\repair\ should exploit repetitions in the sets of documents
where different words belong. It is well known that words do not
distribute uniformly across documents \cite{BYN04}. However, this 
is not the main source of the success of \repair. 
To empirically validate this claim, we modified our inverted
lists as follows: Each list of $\ell$
entries in $[1,u]$ was replaced by $\ell$ different random numbers in $[1,u]$.
So the list lengths are maintained, but the skewness in the chosen documents
is destroyed. \repair\ compressed this new list to 64.24MB, compared to the
48.24MB obtained on the original list.

Figure~\ref{fig:distr} (right) shows the compression ratio
achieved as a function of the length of the lists (dictionary
excluded), for both the real and the random distributions.
The behavior of \repair\ is very similar in both
cases, and thus it can be largely explained by combinatorial
arguments and by the distribution of the list lengths.
This is governed by
Zipf Law \cite{Zip49}. Nevertheless, simple binomial-
 and Poisson/exponential-based models
that ignore interactions between consecutive pairs do not explain
the results well; a more complex model should be used.

The real lists compress 25\% more than the randomly generated
ones. This is related to the parts of the plot where some words of
intermediate frequency compress very well, and corresponds to
positive correlation of word occurrences, that is, words that tend
to appear in the same documents and thus generate the same pairs,
which are thus easily compressed. Therefore, the uneven distribution of words
across the collection \cite{BYN04} is a non-negligible, yet secondary,
source of \repair\ compressibility, being Zipf Law the main one.

Finally, we have experimented with the maximum height of a rule to verify
the hypothesis of logarithmic behavior. We packed 1 to 128 consecutive
documents in the whole collection, so as to have an increasing number of
(aggregated) documents. The growth of the maximum height is indeed logarithmic,
starting from 15 when 128 documents are packed (fewest documents) and
stabilizing around 25 when packing 8 documents. When we optimize the number
of rules, the heights go to 9 and 19 respectively.

\subsection{Time performance}

We now consider intersection time. We show results for two different 
versions of the indexes. In the first scenario we assume that
all the posting lists are compressed with \repair, {\em byte coding}, or \rice\ codes.
In the second scenario, we follow the ideas in \cite{CulpepperM_adcs07} and 
represent the longest lists using bitmaps, and the remainder with the {\em pure} 
techniques.

\subsubsection{{\em Pure compression scenario: byte codes} vs \repair\ vs \rice}

\begin{figure*}[htb]
\centerline{
\includegraphics[width=0.49\textwidth]{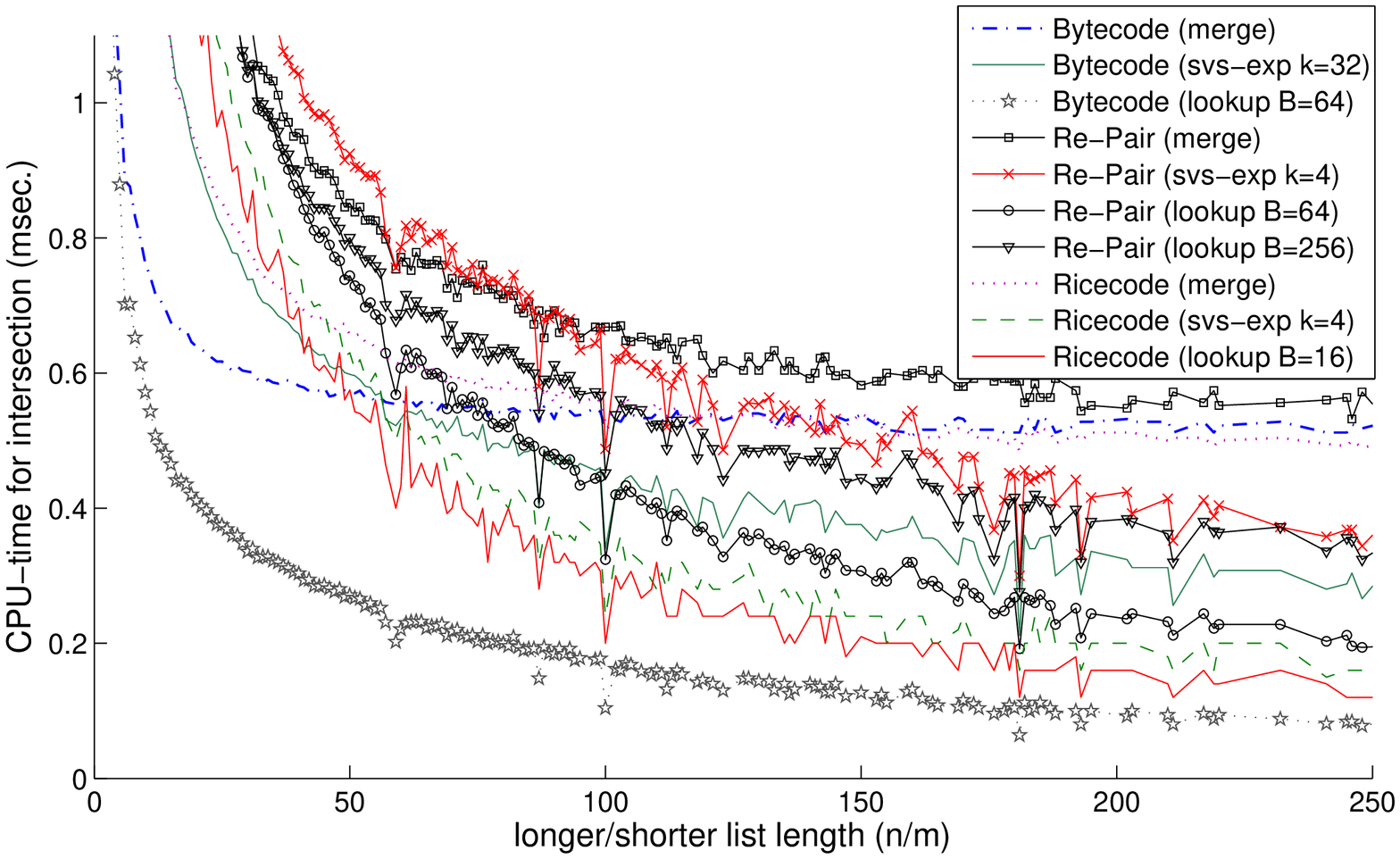}
\includegraphics[width=0.49\textwidth]{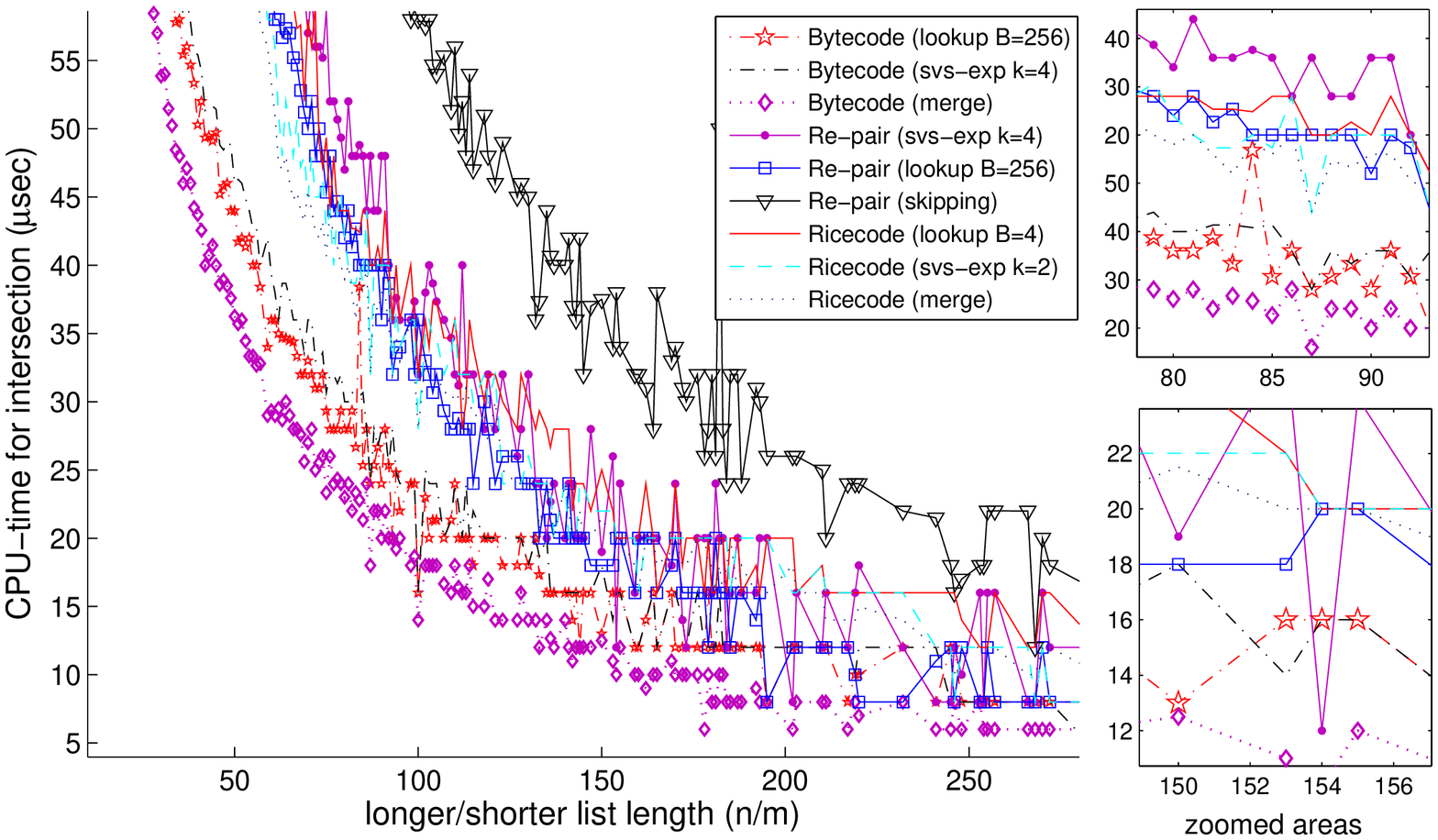}}

\caption{The intersection times as a function of the length ratios
(the longest list has around 100,000 elements). Left: pure variants
for \repair\ and byte codes. Right: variants representing the
longest lists with bitmaps following the approach in
\cite{CulpepperM_adcs07}.} \label{fig:times}
\end{figure*}

\begin{figure}[htb]
\centerline{
\includegraphics[width=0.95\textwidth]{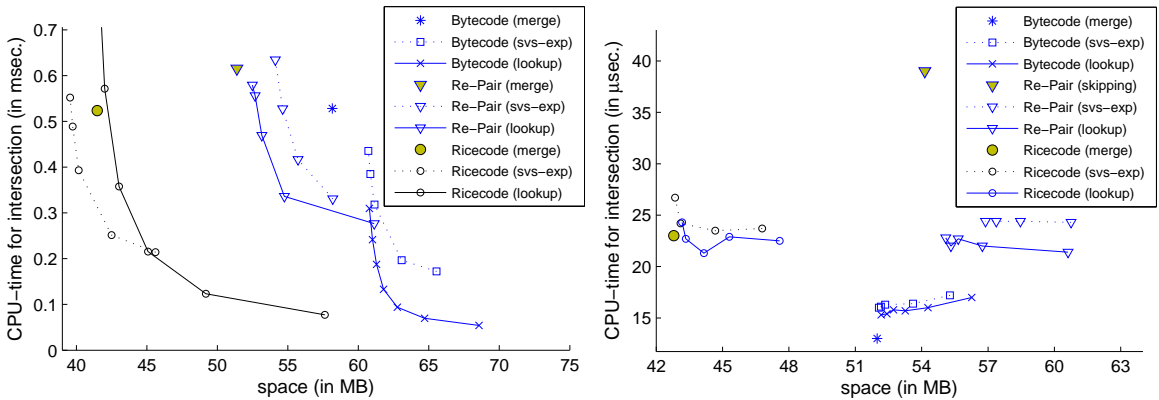}}
\caption{Time-space tradeoff. Intersection times shown as the 
average of runs with $100 \leq n/m \leq 200$. Left: pure variants
for \repair\ and byte codes. Right: variants representing the
longest lists with bitmaps following the approach in
\cite{CulpepperM_adcs07}.}
\label{fig:tradeoff}
\end{figure}

The outcome of the comparison significantly depends on the ratio of lengths
between the two lists \cite{ST07}. Thus we present results between
randomly chosen pairs of words, as a function of this ratio. 
Due to its non-monotonicity, however, the results for \repair\ depend on the absolute length of
the lists. Thus we obtained results for different length ranges of the
longer list. The results given in Figures~\ref{fig:times} and \ref{fig:tradeoff} show times
assuming that the longest list has around 100,000 values. We
generated 1,000 pairs per plot and repeated each search 1,000 times;
average times are computed grouping by ratio.

Figure~\ref{fig:times} (left) shows some results. We chose
variants using least space for byte codes (but they are still much larger than
ours). The \rice\ variants used are clearly the least space demanding 
alternatives. When using \merge, byte codes are faster than \repair, yet
the latter uses significantly less space.

\repair\ with (b)-sampling and \lookup\ search outperforms byte
codes with (a)-sampling and \expon\ search, even using less space
(for $B=64$). Times only match if \repair\ uses $B=256$, but then
\repair\ uses much less space than byte code-based ones. Also,
\repair\ with \expon\ search is a bit slower than with \lookup\ for
about the same space. However, it still performs better than byte
codes using \expon\, for the same space: Although not shown in the
figure, \repair\ with (a)-sampling and $k=1$ requires 58.18MB,
whereas byte coding for $k=32$ needs 60.86MB, yet it performs
similarly to \repair\ with (b)-sampling and $B=64$, which requires
only 54.75MB.

The results worsen a bit for \repair\ on the shorter lists, as \svs\
using \expon\ search obtains similar results to \repair\ with
(b)-sampling and $B=64$, and \merge\ becomes better than \repair\
with no sampling. Again, (b)-sampling is the best choice for
\repair.

Our method dominates the time-space tradeoff when compared to byte
codes with \lookup, yet the latter allows to achieve better times by
letting the structure use much more space than ours. The advantage
of \repair\ can be attributed to the fact that by achieving much
better space, it allows to use a denser sampling, and to the ability
of skipping phrases.

\rice\ coding behave particularly well in this scenario. They require 
much less space than the others (for example \rice\ with (a)-sampling
and $k=4$ requires only 42.49MB), and even though when combined with 
either \merge\ or \lookup\ it is overcome by byte coding (but using 
much less space), it is very competitive when coupled with a \svs\   
intersection algorithm.

Figure~\ref{fig:tradeoff} (left) shows this time-space tradeoff for those
runs with $n/m$ values in the range $100 \leq n/m \leq 200$. 
Our \repair\ with no sampling achieves
13\% better compression than byte codes with no sampling, hence we
can use a denser sampling for the same space (and even
less)\footnote{Note that this sampling is measured over phrases.}.
The dictionary is negligible and it
would fit in RAM for very large collections, even if it scaled
linearly with the data. If we consider the dictionary and the
resulting sequence, the total space is about 10\% of the text size
and, discounting vocabulary, about 25\% of the plain integer
representation of inverted lists. When packing 10 documents into
one, \repair\ total space improves even in relative terms: it is
about 5\% of the text, discounting vocabulary it is 20\% of an
integer inverted list representation, and it is 27\% better than the
byte code-based techniques.

\subsubsection {\em Hybrid compression: bitmaps + pure techniques}

In Figure \ref{fig:times} (right) we include the results representing
the longest lists using bitmaps \cite{CulpepperM_adcs07}. In this
case the other methods outperform \repair\ in almost all aspects.
Figure \ref{fig:tradeoff} (right) shows the tradeoff offered 
by the three methods combined with the bitmap representation
 \cite{CulpepperM_adcs07}. As mentioned above,
the other structures improve further than ours with this approach,
offering a better time/space tradeoff. On the one hand, by representing 
the longest lists with a bitmap, \repair\ cannot benefit from the 
existence of the very repetitive gaps that occur on those lists (it was 
the main source of its good compression). On the other hand, byte coding
 has no longer to pay so much space for representing the longest lists. 
Note that the shortest codeword length is one byte for the byte codes.

\begin{figure}[htb]
\centerline{
\includegraphics[width=0.32\textwidth]{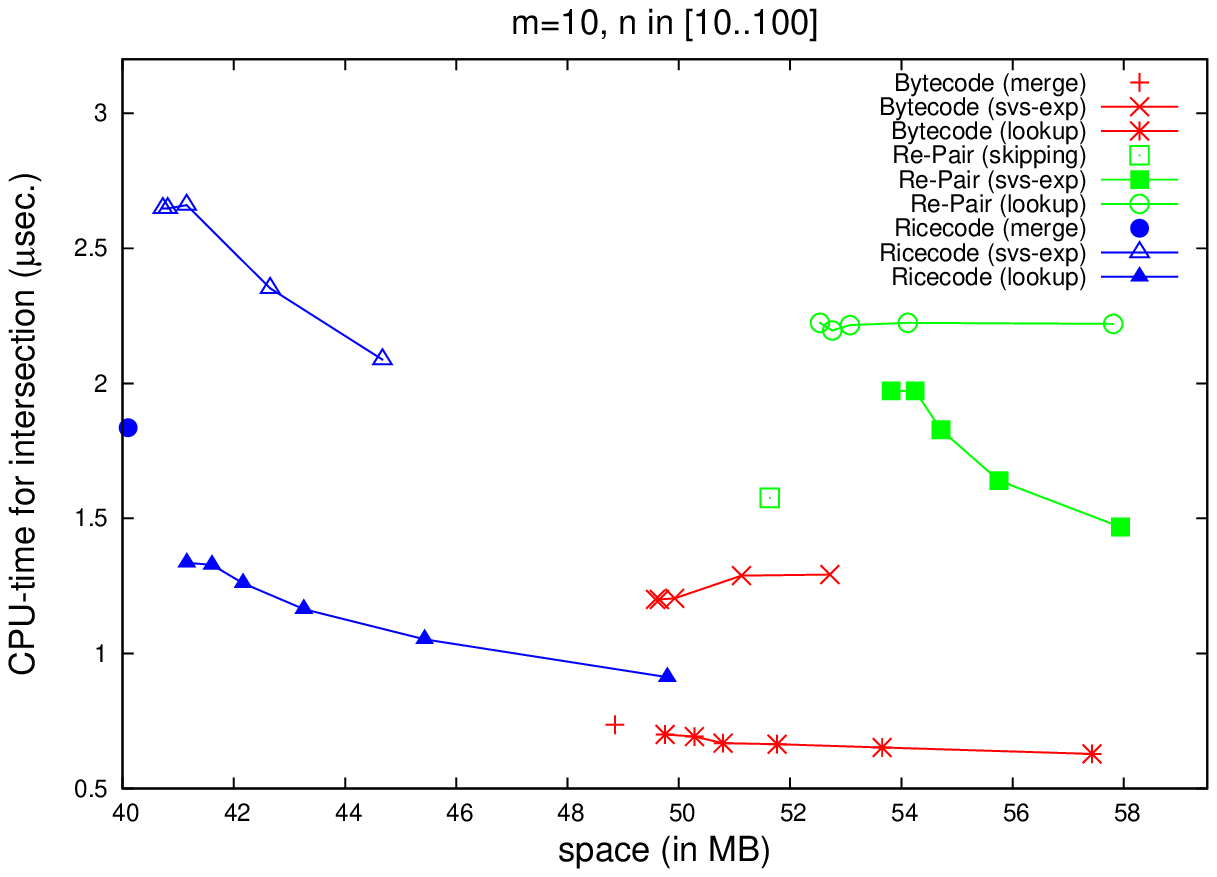}
\includegraphics[width=0.32\textwidth]{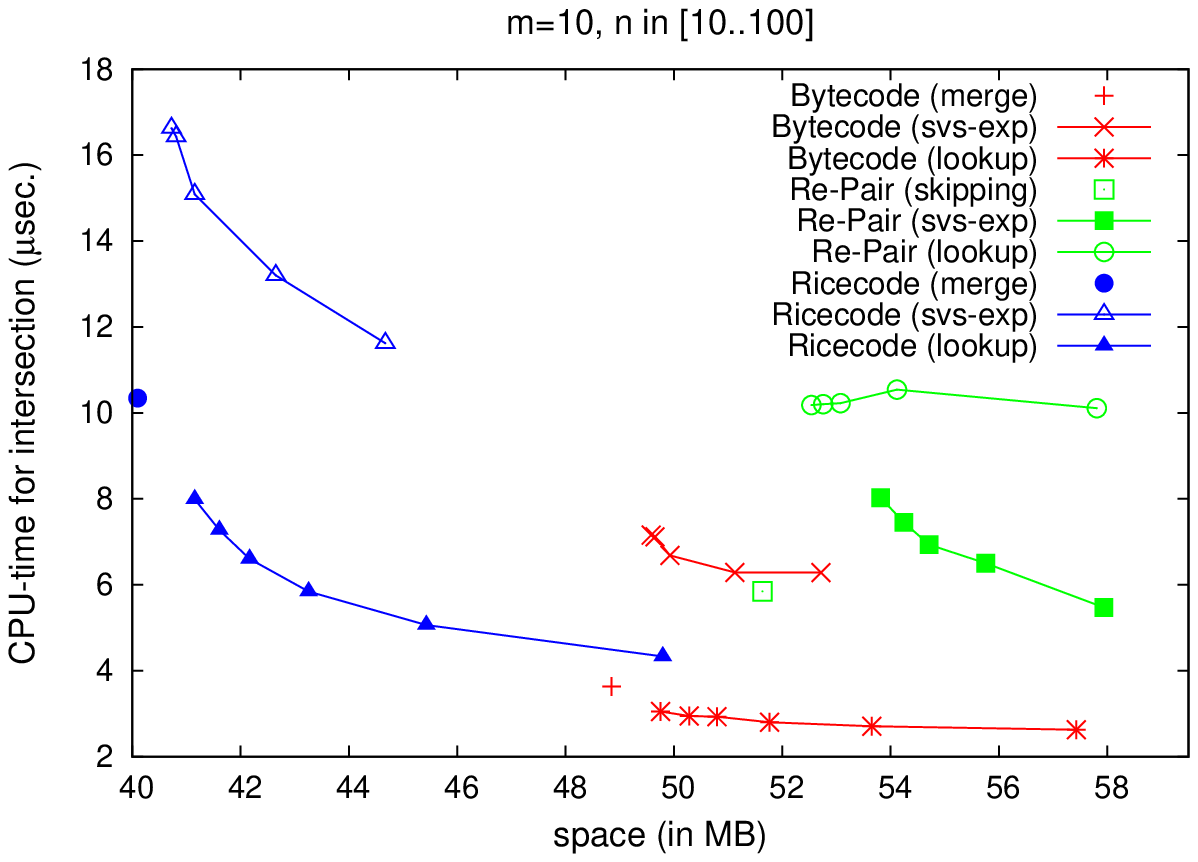}
\includegraphics[width=0.32\textwidth]{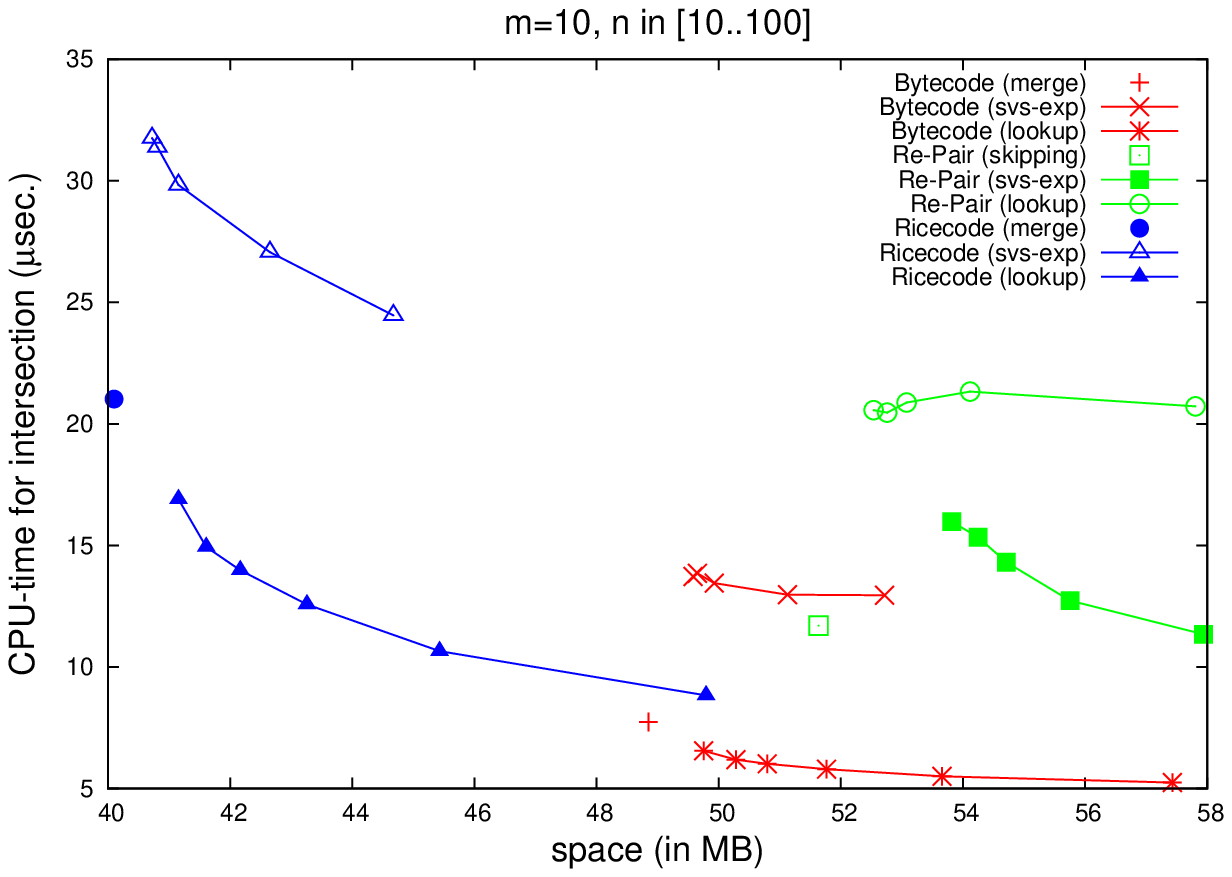}}
~\\
\centerline{
\includegraphics[width=0.32\textwidth]{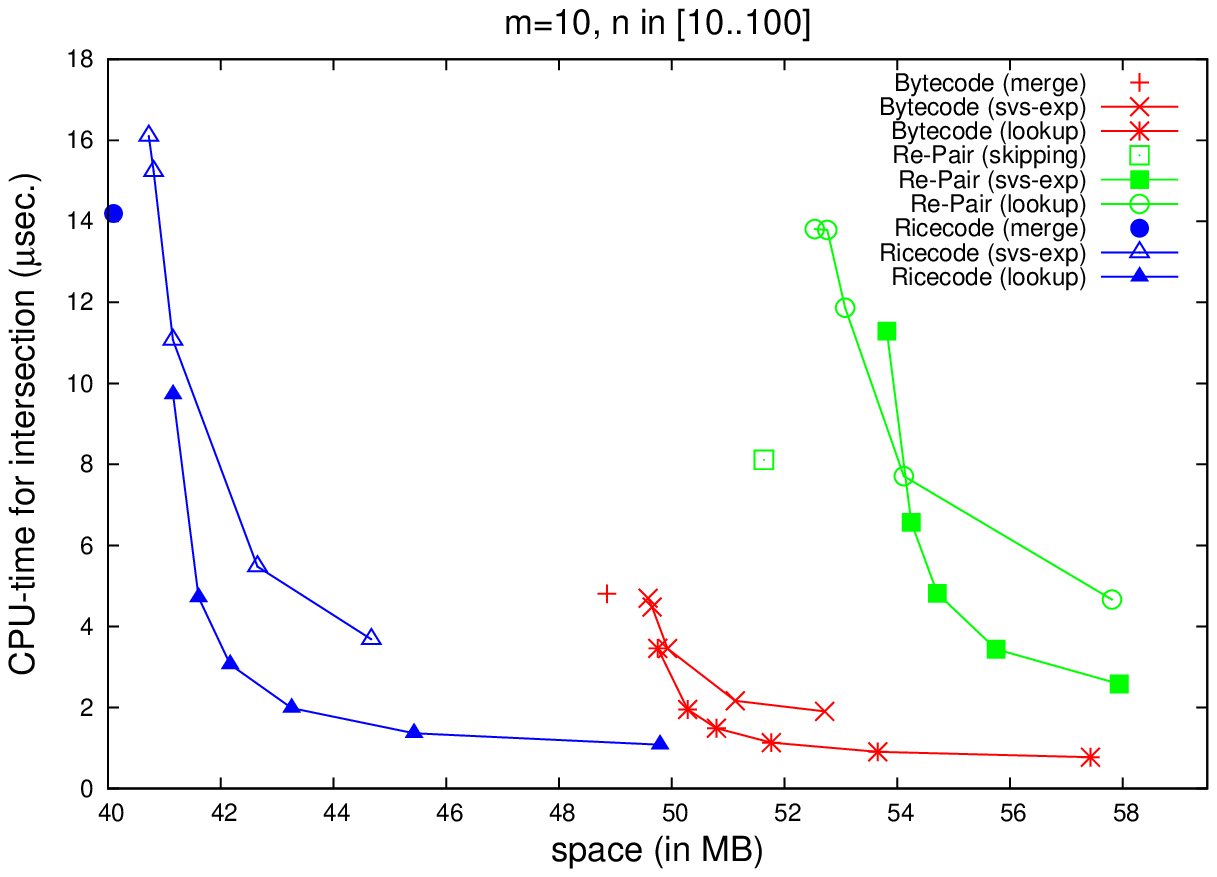}
\includegraphics[width=0.32\textwidth]{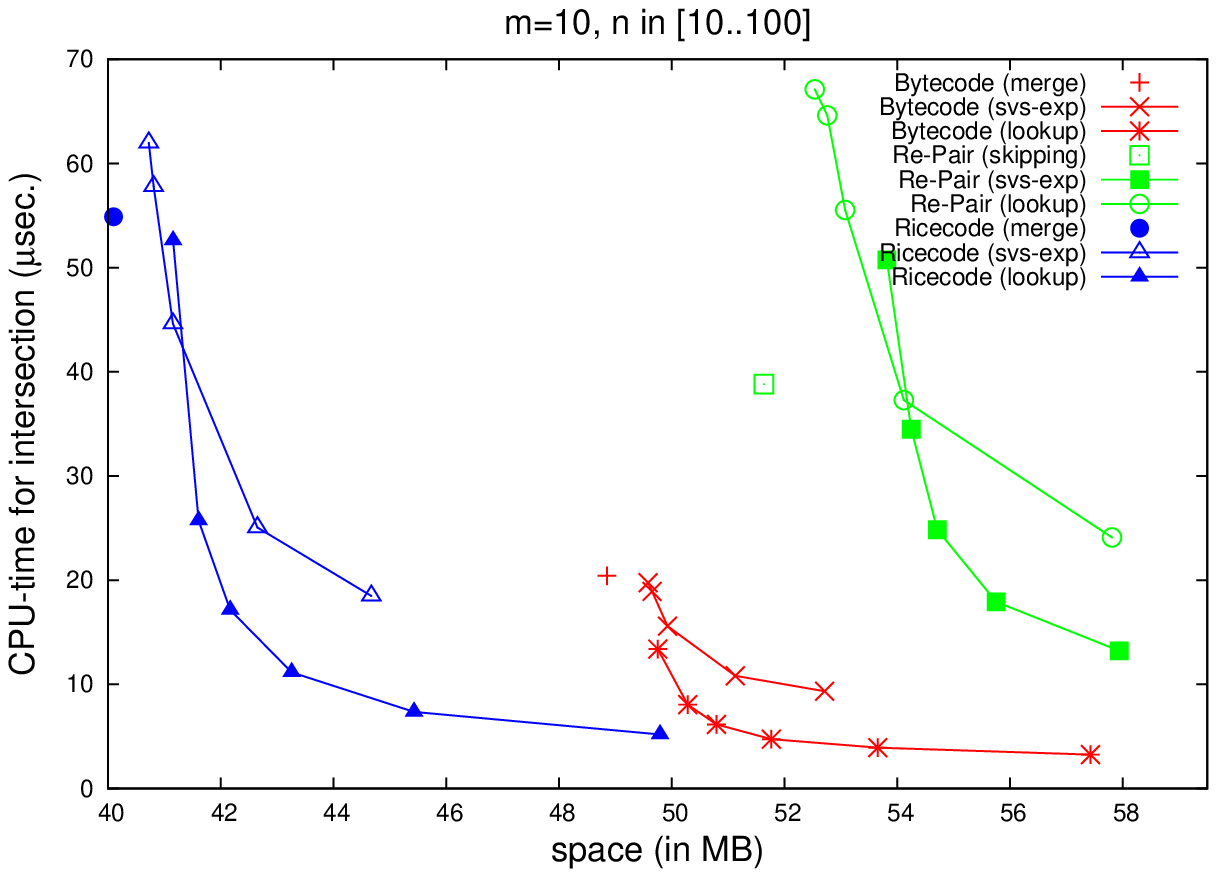}
\includegraphics[width=0.32\textwidth]{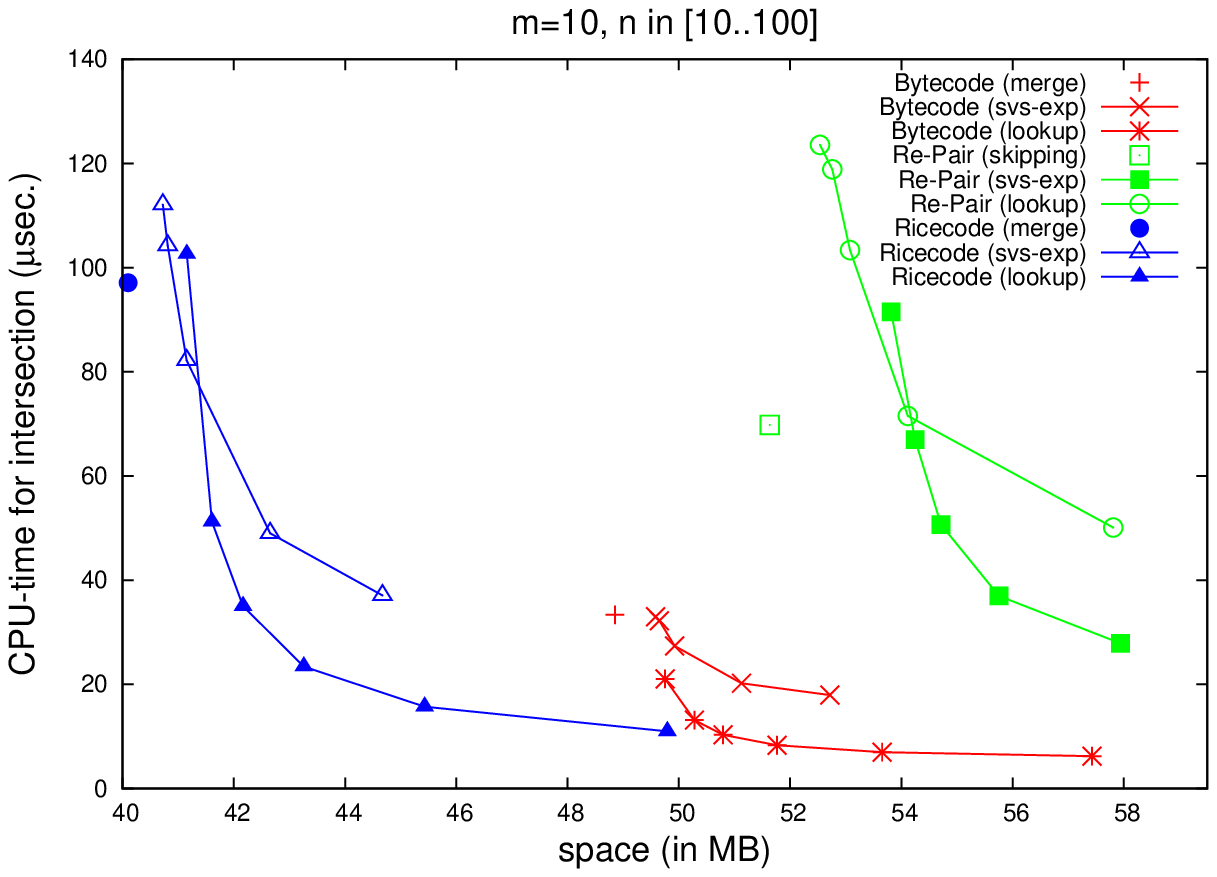}}

\caption{Time-space tradeoff. Intersection times shown as the 
average of runs with $n \in \{10,50,100\}$ and, (left) $n \leq m \leq n\times10$ or (right) $n \leq m \leq n\times100$}.
\label{fig:tradeoff_bitmaps}
\end{figure}

Figure~\ref{fig:tradeoff_bitmaps} shows also values for different lengths of the 
shortest list $n \in \{10,50,100\}$ and the length of longest list $m$ such 
that $n \leq m \leq 10·n$ (first row) and $n \leq m \leq 100·n$ (second row) respectively. 
In these 
experiments, the maximum number of elements in the largest list (n) is always 
$n \leq 10,000$. As proposed in \cite{CulpepperM_adcs07}, we used the value
$ \frac{num\ of\ docs}{8}$ as a threshold for the number of elements a list must contain 
to be compressed with a bitmap. Therefore, we can ensure that no bitmap-compressed
lists will be included in the intersection of two given lists. 
It can be seen that byte-coding is the fastest approach, \rice\
achieves the best compression values, and \repair\ is clearly overcome in 
the space-time tradeoff. The comparison of \merge\ with \rice\ codes
against \repair\ with skipping is more attractive for \repair, as it takes again
advantage of its implicit skipping features.

Given the results in both cases (with and without bitmaps), we
conjecture that the loss of \repair\ in the time/space tradeoff is
due to the fact that \repair\ does not gain as much compression as the other techniques
when converting the longer lists to bitmaps. This suggests that the criteria for
replacing a list and use a bitmap should be studied further.  Our experiments expose a 
result that is of independent interest, namely the {\em lookup} strategy
\cite{ST07} with bitmaps achieves a better time/space tradeoff than 
the original proposal \cite{CulpepperM_adcs07}.

\section{Conclusions}

We have presented a novel method to compress inverted lists. While
previous methods rely on variable-length encoding of differences, we
use \repair\ on the differences. The compression we achieve is not
only much better than difference encoding using byte codes (which
permits denser sampling for the same space), but it also contains
implicit data that allows for fast skipping on the unsampled areas.
Thus our method achieves a good time/space tradeoff in main memory
which, as explained in the Introduction, is a scenario receiving
much attention. When we include the technique proposed by Moffat and
Culpepper \cite{CulpepperM_adcs07}, \repair\ loses its advantage.
Yet, we believe that further research in this line could achieve more
competitive results for our \repair\ representation. For example, 
we could represent with bitmaps the lists that remain long after
compressing.

Because of its locality properties, we also expect our index to perform well on
secondary memory. The vocabulary, the samplings, and the \repair\ dictionary can
realistically fit in RAM: all are small and can be controlled at will. 

We have compared \repair\ with difference encoding using byte codes, as they
have been shown to offer a very good time/compression tradeoff \cite{CM07}.
Byte coding, however, is not the most space-efficient way to encode
differences, as the experiments show in the results for Rice codes.
This variant dominates the time/space tradeoff.
In this aspect, it is interesting that there are also more compact
representations of the \repair\ output \cite{CN08}, which are 
potentially competitive on secondary memory.

Another interesting challenge is how to handle changes to
the collection, where new documents are inserted into or deleted from the
collection. The usual technique for insertions is to index the new documents
and then merge the indexes, so that some symbols are appended at the end of
several lists.

Finally, we also aim at compressed \repair\ dictionary representations that allow
descending faster in the parse tree, and at \repair\ variants for
compression of other types of inverted indexes, such as those used for relevance
ranking.

\section*{Acknoledgements}

This work was funded in part by NSERC of Canada and Go-Bell Scholarships Program 
(first author), by MEC grant TIN2006-15071-C03-03 (second author), by AECI grant 
A/8065/07 (second and third authors), and by Fondecyt grant 1-080019 (first and 
third authors).

\bibliographystyle{alpha}
\bibliography{paper}

\end{document}